\documentclass[sigconf]{acmart}
\usepackage{listings}
\usepackage{algpseudocode}
\usepackage{algorithm}
\newfloat{lstfloat}{htbp}{lop}
\floatname{lstfloat}{Listing}
\setcopyright{none}
\renewcommand\footnotetextcopyrightpermission[1]{}
\usepackage{footnote}
\usepackage{xcolor}
\usepackage{tikz}

\lstdefinelanguage{mlir}{
    alsodigit = {.},
    keywords = {stencil.apply, stencil.access, arith.constant, arith.addf, arith.mulf, stencil.return, f32, f64, stencil.index,stencil.temp}
}

\AtBeginDocument{%
  \providecommand\BibTeX{{%
    \normalfont B\kern-0.5em{\scshape i\kern-0.25em b}\kern-0.8em\TeX}}}

\newcommand\copyrighttext{%
  \footnotesize \textcopyright Nick Brown 2023. This is the author's version of the work. It is posted here for your personal use. Not for redistribution. The definitive version was published in ACM Workshops of The International Conference on High Performance Computing, Network, Storage, and Analysis (SC-W 2023), https://doi.org/10.1145/3624062.3624234"}
\newcommand\copyrightnotice{%
\begin{tikzpicture}[remember picture,overlay]
\node[anchor=south,yshift=10pt] at (current page.south) {\fbox{\parbox{\dimexpr\textwidth-\fboxsep-\fboxrule\relax}{\copyrighttext}}};
\end{tikzpicture}%
}

\begin{document}

\title{Is RISC-V ready for HPC prime-time: Evaluating the 64-core Sophon SG2042 RISC-V CPU}

\author{Nick Brown}
\email{n.brown@ed.ac.uk}
\orcid{0000-0003-2925-7275}
\affiliation{%
  \institution{EPCC at the University of Edinburgh}
  \streetaddress{The Bayes Centre, 47 Potterrow}
  \city{Edinburgh}
  \country{UK}
}
\author{Maurice Jamieson}
\orcid{0000-0003-1626-4871}
\affiliation{%
  \institution{EPCC at the University of Edinburgh}
  \streetaddress{The Bayes Centre, 47 Potterrow}
  \city{Edinburgh}
  \country{UK}
}

\author{Joseph Lee}
\orcid{0000-0002-1648-2740}
\affiliation{%
  \institution{EPCC at the University of Edinburgh}
  \streetaddress{The Bayes Centre, 47 Potterrow}
  \city{Edinburgh}
  \country{UK}
}

\author{Paul Wang}
\orcid{0009-0009-7595-1381}
\affiliation{%
  \institution{PerfXLab}
  \streetaddress{}
  \city{Beijing}
  \country{China}
}


\begin{abstract}
The Sophon SG2042 is the world's first commodity 64-core RISC-V CPU for high performance workloads and an important question is whether the SG2042 has the potential to encourage the HPC community to embrace RISC-V.

In this paper we undertaking a performance exploration of the SG2042 against existing RISC-V hardware and high performance x86 CPUs in use by modern supercomputers. Leveraging the RAJAPerf benchmarking suite, we discover that on average, the SG2042 delivers, per core, between five and ten times the performance compared to the nearest widely available RISC-V hardware. We found that, on average, the x86 high performance CPUs under test outperform the SG2042 by between four and eight times for multi-threaded workloads, although some individual kernels do perform faster on the SG2042. The result of this work is a performance study that not only contrasts this new RISC-V CPU against existing technologies, but furthermore shares performance best practice.
\end{abstract}



\maketitle
\pagestyle{plain}
\copyrightnotice
\section{Introduction}
In recent years RISC-V has become a well-established open ISA standard with over 10 billion RISC-V CPU cores having been manufactured at the time of writing. Whilst RISC-V has gained traction in embedded computing fields such as automotive, space, and micro-controllers it can also potentially offer benefits to a variety of other areas of computing including High Performance Computing (HPC).

Whilst the HPC community moves further into the exascale era, an important question is what technologies will power our future supercomputers. To this end there have been numerous RISC-V software activities focused on HPC, however a major challenge has been the availability of physical hardware which is capable of supporting such workloads. There are several RISC-V HPC hardware efforts in development, but apart from a few prototypes the HPC community has only been able to obtain widespread access to embedded Single Board Computers (SBC), often with more of a focus on energy efficiency and small core counts than high performance. Consequently, whilst some preparatory work is being undertaken around RISC-V for HPC \cite{mantovani2023software} \cite{herdt2020risc} \cite{perez2021coyote}, it is difficult to encourage the HPC community to fully embrace the potential of RISC-V without serious hardware propositions.

However, RISC-V is moving very rapidly and in the summer of 2023 the 64-core Sophon SG2042 CPU was released. This is the world's first publicly available, mass produced, 64-core RISC-V CPU aimed at high performance workloads. Not only does this processor provide a step change in terms of the number of cores available, but furthermore these cores are T-Head XuanTie C920s which a marketed as being designed for high performance. Consequently this new RISC-V CPU is very interesting to the HPC community and has the potential to bring RISC-V to high performance workloads. In this paper we benchmark the SG2042 using the popular RAJAPerf HPC performance suite. Comparing this CPU to existing RISC-V processors to understand whether the promised step change in capability is delivered compared to other RISC-V commodity options. We then explore how configuration and tool choices impact the performance of workloads on the SG2042 before comparing this 64 core CPU against high performance x86 CPUs that are commonly used in supercomputers. Our ultimate aim is to answer the central question of whether the SG2042 means that RISC-V can now be considered a more serious proposition for HPC workloads. 

This paper is structured as follows; the background to this work is described in Section \ref{sec:bg} where we present technical details of the Sophon SG2042 CPU and host machine used in this work as well as the RAJAPerf benchmark suite. Section \ref{sec:benchmarking} then reports the results of our benchmarking activities, first describing comparisons against existing RISC-V CPUs in Section \ref{sec:riscv-comparison}, then exploring the impact of configurations and tools on performance in Section \ref{sec:perf_exploration} to better understand best practice, before Section \ref{sec:x86_performance} compares against high performance x86 CPUs common in HPC machines. Lastly, Section \ref{sec:conc} draws conclusions and discusses further work.

It should be stressed that this is an independent benchmarking study of the Sophon SG2042 and the authors of this work have no links to the manufacturers of the CPU or cores. To the best of our knowledge this is the world's first independent benchmarking study of a high performance 64-core RISC-V CPU.

\section{Background}
\label{sec:bg}
In early 2023, RISC-V International identified HPC as a strategic priority area for growth and from this, combined with activities such as the recently ratified vector extension and numerous HPC software efforts which are porting key HPC libraries and tools, it is evident that momentum in this space is growing rapidly. Activities around the world, such as the European eProcessor project \cite{eprocessor}, the thousand core Esperanto CPUs \cite{esperanto}, and the multi-vendor RISE project \cite{eprocessor} which aims to develop critical software component support for RISC-V, have the potential to popularise RISC-V in high-end computing, including HPC, and ultimately enable the community to build supercomputers around this technology. Furthermore, early application studies report favourably on the benefits that RISC-V can deliver to high performance workloads.

However, whilst there are numerous companies working on prototype high performance RISC-V hardware, to date choice has been extremely limited when looking to run workloads on commodity available RISC-V hardware. There has been an unfortunate choice between either using soft-cores\footnote{Soft-cores are software descriptions of a CPU that run on an FPGA, however their clock frequency tends to be far lower than a physical CPU} or physical cores designed more for energy efficient and embedded workloads. Irrespective, whilst these solutions enable experimentation with RISC-V, they do not provide the capabilities required for production high performance workloads on the architecture. Consequently, whilst there is interest in the HPC community around RISC-V, it is yet to embrace the technology.

\subsection{The Sophon SG2042 CPU}
\label{sec:cpu_desc}
The Sophon SG2042 CPU is a 64-core processor running at 2GHz and organised in clusters of four XuanTie C920 cores. Each 64-bit core, designed by T-Head, is designed for high performance and adopts a 12-stage out-of-order multiple issue superscalar pipeline design. Providing the RV64GCV instruction set, the C920 has three decode, four rename/dispatch, eight issue/execute and two load/store execution units. Version 0.7.1 of the vectorisation standard extension (RVV v0.7.1) is supported, with a vector width of 128 bits supporting data types FP16, FP32, FP64, INT8, INT16, INT32, and INT64. Each C920 core contains 64KB of L1 instruction (I) and data (D) cache, 1MB of L2 cache which is shared between the cluster of four cores, and 64MB of L3 system cache which is shared by all cores in the package. The SG2042 also provides four DDR4-3200 memory controllers, and 32 lanes of PCI-E Gen4. The CPU we use for the benchmarking in this paper is mounted in a Pioneer Box by Milk-V which contains 32GB of RAM and a 1TB SSD. 

An important consideration for HPC workloads is that of vectorisation and due to the C920 cores only supporting RVV v0.7.1, compiler support is a challenge. The current upstream version of the RISC-V GNU compiler does not provide support for any version of the vector extension. Whilst the GNU repository contains an \emph{rvv-next} branch \cite{rvv-next} whose purpose is to support RVV v1.0, at the time of writing this is not actively maintained. Furthermore there was an \emph{rvv-0.7.1} branch which targeted RVV v0.7.1 but this has been deleted. Due to this lack of support in mainline GCC, T-Head, the chip division of Alibaba, have provided their own fork of the GNU compiler (XuanTie GCC) which has been optimised for their processors.

T-Head's bespoke compiler supports both RVV v0.7.1 and their own bespoke custom extensions. Whilst several versions of this compiler have been provided, it has been found \cite{lee2023backporting} that GCC8.4 which is part of their 20210618 release provides the best auto-vectorisation capability and-so this is the version we have selected for the benchmarking experiments undertaken in this paper. Their version of the compiler generates Vector Length Specific (VLS) RVV assembly which specifically targets the 128-bit vector width of the C920. All kernels are compiled at optimisation level three, and all reported results are averaged over five runs. All benchmarks are executed directly from the Linux command line rather than, for example, being submitted via a job scheduling system. At the time of execution each benchmark run was making exclusive use of the machine.

\subsection{RAJAPerf}
\label{sec:raja}
The RAJA performance suite \cite{hornung2017raja} is designed to explore the performance of loop-based computational kernels, which are common in HPC applications. Whilst it was initially developed as a tool to benchmark the performance of the RAJA parallel programming framework \cite{beckingsale2019raja}, the suite has been extended and developed to support a variety of different targets, such as OpenMP. This popular benchmarking suite has been used extensively for testing HPC hardware and tools, and consequently RAJAPerf has been ported into numerous languages \cite{jesus2022chapelperf} and used for a variety of studies \cite{jesus2023check} \cite{jost2021seamless}.

The benchmark comprises of 64 kernels which are categorised into six classes:

\begin{itemize}
  \item \textbf{Algorithm}: Contains six kernels which undertake basic algorithmic activities such as memory copies, the sorting of data and reductions.
  \item \textbf{Apps}: Comprises thirteen kernels and these represent common components of HPC applications such as an FIR filter, data packing and unpacking for halo exchanges, 3D diffusion and convection by partial assembly, and solving LaPlace's equation for diffusion in 1D.
  \item \textbf{Basic}: Represents foundational mathematical functions via sixteen kernels. These include DAXPY, matrix multiplication, integer reduction, and calculation of PI by reduction.
  \item \textbf{Lcals}: The Livermore Compiler Analysis Loop Suite which is a collection of eleven loop based kernels including tridiagonal elimination, calculation of differences, and calculations of minimums and maximums.
  \item \textbf{Polybench}: Thirteen polyhedral kernels which includes two and three matrix multiplications, matrix transposition and vector multiplication, a 2D Jacobi stencil computation, and an alternating direction implicit solver.
  \item \textbf{Stream}: Five kernels that focus on memory bandwidth and the corresponding computation, these are based upon simple vectorisable functions. 
\end{itemize}

Given the differences between these classes, it is interesting to explore how different features of the hardware or software under test perform with specific aspects. In a previous study \cite{lee2023backporting}, RAJAPerf was used to compare the performance of the four core SiFive VisionFive V2 against the single core AllWinner D1. Whilst these processors are designed for embedded workloads, the AllWinner D1 contains the XuanTie C906 core which, although it is designed for energy efficiency rather than performance \cite{c906}, provides support for the RVV v0.7.1 extension. In that study it was found that whilst the U74 core in the VisionFive V2 tended to outperform the C906 for scalar workloads, when enabling vectorisation the C906 then most often outperformed the U74. However, these machines provide small core count CPUs containing cores which are not designed for high performance workloads. Consequently, whilst \cite{lee2023backporting} was an interesting study, the hardware under test was never a realistic option for production HPC workloads.

\section{Benchmarking}
\label{sec:benchmarking}
\subsection{Comparison against RISC-V processors}
\label{sec:riscv-comparison}
In this section we compare performance of the Sophon SG2042 against a StarFive VisionFive V1 and StarFive VisionFive V2. The V1 contains the JH7100 SoC, whereas the V2 contains the JH7110. Both these SoCs are built around the 64-bit RISC-V SiFive U74 core, with the JH7100 containing two and the JH7110 containing four cores. The JH7100 is running at 1.2GHz and the JH7110 at 1.5Ghz \cite{starfive-soc}, with the U74 cores contain 32KB D and 32KB I L1 cache and both SoC models also containing 2MB of L2 cache shared between the cores. However, only RV64GC is provided by the SiFive U74, and consequently there is no support for the RISC-V vector extension.

\begin{figure}[htb]
\centering
 \includegraphics[width=\columnwidth]{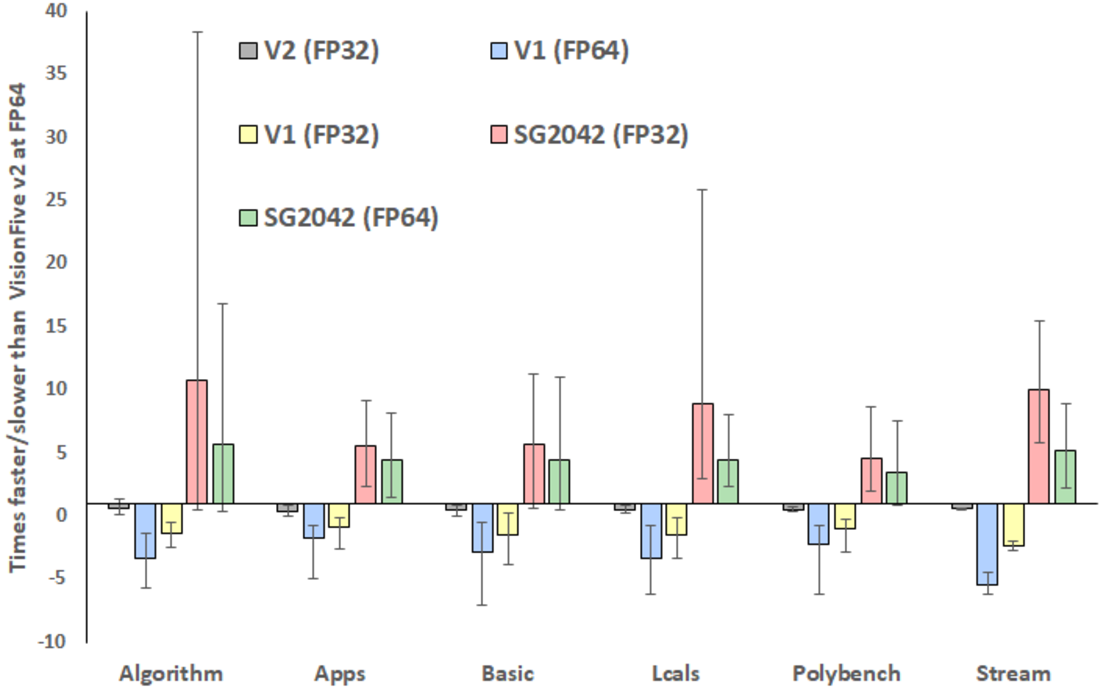}
\caption{Single core comparison baselined against StarFive VisionFive V2 running in double precision (FP64), against V1 and SG2042}	
\label{fig:riscv-comparison}
\end{figure}

Figure \ref{fig:riscv-comparison} reports a performance comparison of the VisionFive V2 and V1 against the SG2042 for double (FP64) and single (FP32) precision. The numbers reported are relative to the performance of the V2 running the RAJAPerf benchmark suite at double precision as a baseline. Zero on the graph indicates the same performance, positive numbers are the number of times that the configuration is faster than the baseline, and negative numbers is the number of times slower. The bars in the graph report an average across the specific RAJAPerf class, as described in Section \ref{sec:raja}, and the whiskers report the max-min range, with the top of the whisker being the maximum speedup compared to the baseline and the bottom of the whisker the minimum speedup (or maximum slow down).

It can be seen from Figure \ref{fig:riscv-comparison} that a single C920 core of the SG2042 outperforms the U74 core of the V2 and V1 at both double and single precision. At double precision the C920 core delivers on average between 4.3 and 6.5 times the performance achieved when running at double precision on the U74 in the V2. Furthermore, at single precision the C920 achieves between 5.6 and 11.8 times the performance on average across the benchmarks. This is an impressive performance gain, and there were no kernels that ran slower on the C920 core than the U74. The performance of some kernels was very impressive on the C920, for instance the memory set benchmark from the algorithm group ran 40 times faster in FP32 and 18 times faster in FP64 than on the U74.

It should be highlighted that we are running the benchmarks on these cores at their best possible configuration. Effectively this means that we leverage vectorisation on the SG2042's C920, whereas this is not supported by the U74 and hence unavailable on the V1 or V2. The results in Figure \ref{fig:riscv-comparison} demonstrate a noteworthy performance difference between FP32 and FP64 on the SG2042. In part, this is likely due to the width of the vector registers on the SG2042, which are 128 bits, and-so four floating point numbers can be operated upon with FP32, whereas only two with FP64. By comparison, the performance difference between running double and single precision on the V2 is far less but the V2 does not support RVV.

An aspect of the results in Figure \ref{fig:riscv-comparison} that surprised us is how much slower the VisionFive V1 is than the V2. Considering that we are running RAJAPerf over a single core only, so the dual vs quad core nature of these machines does not matter, and that they both contain the same U74 core, even though the V1 is running at 1.2GHz and the V2 at 1.5GHz, one would assume that performance would be relatively similar. However, at double precision the V1 is between six and three times slower than the V2, and at single precision between one and three times slower. Whilst the lower clock frequency of the V1 will impact performance somewhat compared to the V2, this does not explain such a large difference but as a performance comparison between the V1 and V2 is not the objective of this paper, we leave the endeavour of explaining this phenomena to future work.

It can be seen from Figure \ref{fig:riscv-comparison} that the performance obtained by a single C920 core of the SG2042 is impressive compared to existing, publicly available, commodity  RISC-V hardware. This core is described by T-Head as a high performance RISC-V processor, and the benchmarking results reported in this section demonstrate that it delivers a large improvement in performance across the entire benchmarking suite against the U74 which would previously have been considered the best choice of widely available RISC-V CPU to experiment with HPC workloads upon. 

\subsection{SG2042 performance exploration}
\label{sec:perf_exploration}
In Section \ref{sec:riscv-comparison} we compared the SG2042's XuanTie C920 core against other RISC-V commodity hardware that, until now, might be seen as the best choice to experiment upon with high performance workloads. However, in addition to single core performance the SG2042 is also significantly ahead of the V1's JH7100 and V2's JH7110 SoCs in terms of the number of cores. Consequently, it is interesting to explore this facet in more detail to understand the performance properties at different configurations for programmer best practice, and in this section we also explore the vectorisation of the C920 core too.

\begin{table*}[ht]
\begin{center}
\begin{tabular}{|c|cc|cc|cc|cc|cc|cc|}
\hline
&
\multicolumn{2}{c|}{\textbf{Algorithm}}
& 
\multicolumn{2}{c|}{\textbf{Apps}}
& 
\multicolumn{2}{c|}{\textbf{Basic}}
& 
\multicolumn{2}{c|}{\textbf{Lcals}}
& 
\multicolumn{2}{c|}{\textbf{Polybench}}
& 
\multicolumn{2}{c|}{\textbf{Stream}}
\\

\cline{2-13}

\textbf{Threads} & \textbf{Speedup} & \textbf{PE} 
& \textbf{Speedup} & \textbf{PE} & \textbf{Speedup} & \textbf{PE}
& \textbf{Speedup} & \textbf{PE} & \textbf{Speedup} & \textbf{PE} 
& \textbf{Speedup} & \textbf{PE}\\
\hline
2 & 1.19 & 0.60 & 0.66 & 0.33 & 1.02 & 0.51 & 1.61 & 0.81 & 1.86 & 0.93 & 1.00 & 0.50\\
4 & 1.12 & 0.28 & 1.14 & 0.29 & 1.81 & 0.45 & 1.82 & 0.45 & 3.46 & 0.86 & 0.97 & 0.24\\
8 & 2.02 & 0.25 & 2.27 & 0.28 & 3.55 & 0.44 & 3.27 & 0.41 & 7.72 & 0.96 & 1.88 & 0.24\\
16 & 4.64 & 0.29 & 4.31 & 0.27 & 6.92 & 0.43 & 6.86 & 0.43 & 15.39 & 0.96 & 4.31 & 0.27\\
32 & 1.11 & 0.03 & 1.86 & 0.06 & 0.22 & 0.007 & 4.38 & 0.14 & 14.09 & 0.44 & 0.82 & 0.03\\
64 & 0.97 & 0.02 & 4.10 & 0.06 & 12.33 & 0.19 & 14.89 & 0.23 & 40.42 & 0.63 & 1.77 & 0.03\\

\hline
\end{tabular}
\caption{Speed up and parallel efficiency for benchmark classes as we scale the number of threads, using block allocation of threads to cores where threads map contiguously to CPU cores}
\label{tab:multithread_performance}
\end{center}
\end{table*}

Table \ref{tab:multithread_performance} reports speed up \footnote{Speed up is the execution time on one thread divided by execution on \emph{n} threads} and parallel efficiency \footnote{Parallel efficiency is the speed up divided by the number of threads. This ranges from 1 to 0, where 1 is optimal} obtained across the classes of the RAJAPerf benchmark suite when scaling the number of threads. Throughout our runs we set the \emph{OMP\_PROC\_BIND} environment variable to be true to ensure that threads can not migrate during execution. These multi-threaded runs are undertaken in single precision, FP32.

In the experiment reported in Table \ref{tab:multithread_performance} we assigned threads to cores contiguously in a block allocation approach. For instance, thread one is mapped to core one, thread two mapped to core two, and thread three mapped to core three. It can be seen from these results that although threading is generally beneficial at smaller thread counts, the \emph{apps} class ran slower with two threads compared with one. However, as the thread count increases then the parallel efficiency decreases significantly for some benchmark classes, and in some cases execution time is actually greater on 32 or 64 threads than on one. 

\begin{table*}[ht]
\begin{center}
\begin{tabular}{|c|cc|cc|cc|cc|cc|cc|}
\hline
&
\multicolumn{2}{c|}{\textbf{Algorithm}}
& 
\multicolumn{2}{c|}{\textbf{Apps}}
& 
\multicolumn{2}{c|}{\textbf{Basic}}
& 
\multicolumn{2}{c|}{\textbf{Lcals}}
& 
\multicolumn{2}{c|}{\textbf{Polybench}}
& 
\multicolumn{2}{c|}{\textbf{Stream}}
\\

\cline{2-13}

\textbf{Threads} & \textbf{Speedup} & \textbf{PE} 
& \textbf{Speedup} & \textbf{PE} & \textbf{Speedup} & \textbf{PE}
& \textbf{Speedup} & \textbf{PE} & \textbf{Speedup} & \textbf{PE} 
& \textbf{Speedup} & \textbf{PE}\\
\hline
2 & 1.52 & 0.76 & 0.70 & 0.35 & 1.06 & 0.53 & 1.81 & 0.91 & 2.11 & 1.06 & 1.93 & 0.96\\
4 & 3.21 & 0.80 & 1.37 & 0.34 & 2.09 & 0.52 & 3.61 & 0.90 & 4.11 & 1.03 & 4.19 & 1.05\\
8 & 4.72 & 0.59 & 2.64 & 0.33 & 3.96 & 0.49 & 6.08 & 0.76 & 8.15 & 1.02 & 4.46 & 0.56\\
16 & 4.55 & 0.28 & 4.32 & 0.27 & 6.97 & 0.44 & 7.12 & 0.45 & 15.07 & 0.94 & 4.19 & 0.26\\
32 & 6.10 & 0.19 & 6.32 & 0.20 & 13.11 & 0.41 & 14.84 & 0.46 & 30.05 & 0.94 & 13.91 & 0.43\\
64 & 2.09 & 0.03 & 4.31 & 0.07 & 17.29 & 0.27 & 26.53 & 0.41 & 57.93 & 0.91 & 1.62 & 0.03\\

\hline
\end{tabular}
\caption{Speed up and parallel efficiency for benchmark classes as we scale the number of threads, using cyclic allocation of threads to cores where threads cycle round NUMA regions and are then allocated contiguously in a region}
\label{tab:multithread_performance_cyclic}
\end{center}
\end{table*}

To explain some of the lacklustre scaling seen in Table \ref{tab:multithread_performance} it was our hypothesis that the thread placement we had adopted was causing many of the performance issues. This is because there is one DDR memory controller per NUMA region, and-so it was likely that a large contributor to this poor performance was a bottleneck on individual controllers because our threads were filling up each NUMA region contiguously, thus some NUMA regions contained a large number of active threads whereas others potentially none for medium thread counts. Using the \emph{lscpu} tool, we discovered that the SG2042 contains four NUMA regions. Unusually for physical CPUs (i.e. not SMT), the core ids are not contiguous in a NUMA region but instead eight consecutive cores reside in a NUMA region, then there is a gap of eight and the following eight are also in the NUMA region. Consequently, cores 0-7 and 16-23 are in NUMA region 0, 8-15 and 24-31 are in NUMA region 1, 32-39 and 48-55 are in NUMA region 2, and 40-47 and 56-63 are in NUMA region 3.  

We therefore we ran an experiment which involved allocating threads cyclically across the NUMA regions. For example, four threads are mapped to cores 0, 8, 32, and 40. Beyond this number of threads the placement within a NUMA region is contiguous, for instance eight threads are placed onto cores 0, 8, 32, 40, 1, 9, 33, and 41. Table \ref{tab:multithread_performance_cyclic} reports the speed up and parallel efficiency results with this cyclic placement and it can be seen, compared to the block allocation that was used for experiments in Table \ref{tab:multithread_performance}, this placement policy delivers significantly improved scaling in general. Furthermore, it can be seen that at 64 threads the cyclic allocation policy outperforms the block policy, apart from with the \emph{stream} class, which is surprising as one would assume that because all the cores are allocated then this cycling across the NUMA regions would cease to be beneficial.

\begin{table*}[ht]
\begin{center}
\begin{tabular}{|c|cc|cc|cc|cc|cc|cc|}
\hline
&
\multicolumn{2}{c|}{\textbf{Algorithm}}
& 
\multicolumn{2}{c|}{\textbf{Apps}}
& 
\multicolumn{2}{c|}{\textbf{Basic}}
& 
\multicolumn{2}{c|}{\textbf{Lcals}}
& 
\multicolumn{2}{c|}{\textbf{Polybench}}
& 
\multicolumn{2}{c|}{\textbf{Stream}}
\\

\cline{2-13}

\textbf{Threads} & \textbf{Speedup} & \textbf{PE} 
& \textbf{Speedup} & \textbf{PE} & \textbf{Speedup} & \textbf{PE}
& \textbf{Speedup} & \textbf{PE} & \textbf{Speedup} & \textbf{PE} 
& \textbf{Speedup} & \textbf{PE}\\
\hline
2 & 1.52 & 0.76 & 0.70 & 0.35 & 1.06 & 0.53 & 1.81 & 0.91 & 2.11 & 1.06 & 1.93 & 0.96\\
4 & 3.21 & 0.80 & 1.37 & 0.34 & 2.09 & 0.52 & 3.61 & 0.90 & 4.11 & 1.03 & 4.19 & 1.05\\
8 & 6.37 & 0.80 & 2.71 & 0.34 & 4.16 & 0.52 & 7.15 & 0.89 & 8.23 & 1.03 & 11.20 & 1.40\\
16 & 10.54 & 0.66 & 5.13 & 0.32 & 8.09 & 0.51 & 13.55 & 0.85 & 16.51 & 1.03 & 11.60 & 0.73\\
32 & 12.72 & 0.40 & 8.77 & 0.27 & 14.05 & 0.44 & 21.29 & 0.67 & 31.76 & 0.99 & 15.18 & 0.47\\
64 & 1.98 & 0.03 & 3.69 & 0.06 & 17.30 & 0.27 & 17.70 & 0.28 & 58.26 & 0.91 & 1.51 & 0.02\\

\hline
\end{tabular}
\caption{Speed up and parallel efficiency for benchmark classes as we scale the number of threads, using cluster aware cyclic allocation of threads to cores where threads cycle round NUMA regions and cycle round inside each NUMA region cyclically across the clusters of four cores}
\label{tab:multithread_performance_cluster}
\end{center}
\end{table*}

As described in Section \ref{sec:cpu_desc}, the SG2042's C920 cores are organised in clusters of four with each cluster containing 1MB of shared L2 cache. Consequently, we ran another placement experiment to explore whether, within a NUMA region, also working cyclically across these clusters would provide improved performance as doing so could make better use of the shared L2 cache for smaller to medium thread counts. For example, using this placement then 8 threads would be mapped to cores 0, 8, 32, 40, 16, 24, 48, and 56. 

Table \ref{tab:multithread_performance_cluster} reports the speed up and parallel efficiency for this \emph{cluster} placement policy, and it can be seen that up to and including 32 threads such a policy delivers a noticeable improvement compared to the previous cyclic policy of placing cyclically across NUMA regions but contiguously within a region. At 64 threads, cyclic placement tends to be more beneficial however, as all the CPU cores are active then the benefit of this cluster aware placement will be limited because the L2 cache will be shared between four active cores regardless. Therefore,  based upon these experiments, for performance, the importance of considering the NUMA regions and four core clusters when adopting a thread placement has been demonstrated.

\begin{figure}[htb]
\centering
 \includegraphics[width=\columnwidth]{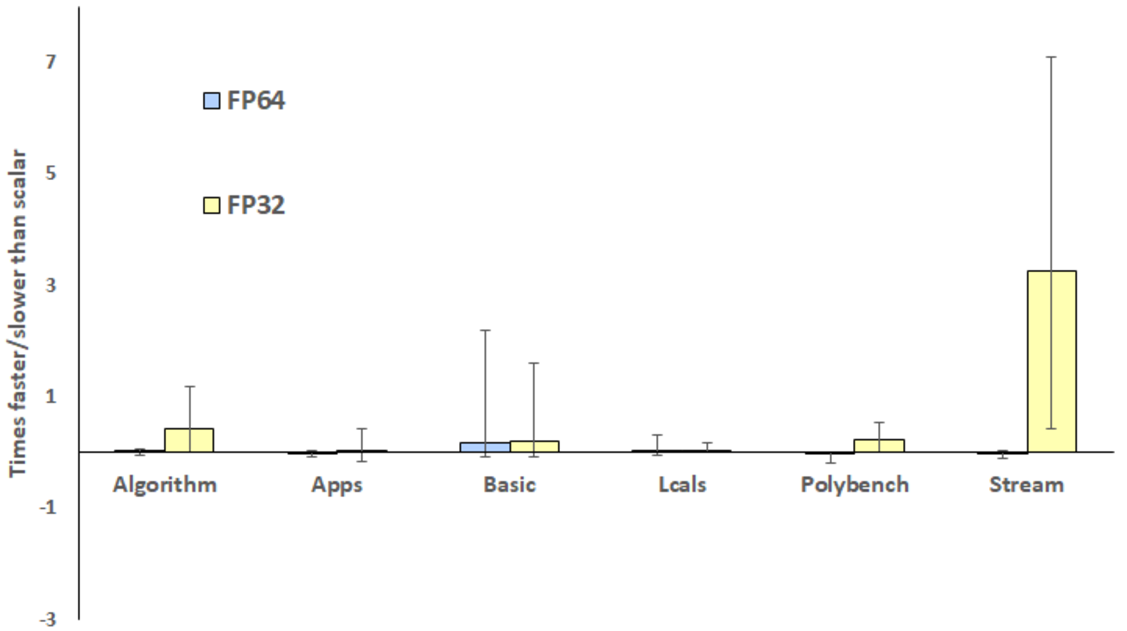}
\caption{Maximum single core speedup for each benchmark class when enabling vectorisation on C920 of SG2042}	
\label{fig:vect-scalar}
\end{figure}

When considering the performance that the SG2042 will deliver, another important consideration is whether to enable vectorisation or not. Figure \ref{fig:vect-scalar} reports the difference in performance, when running on a single core, enabling vectorisation for FP32 and FP64 compared to each of these configurations running scalar-only. The bars depict the average across the benchmark class, with the top of the whiskers being the greatest speed up and the bottom the lowest. Zero means that performance is the same, one means that performance is one time faster (e.g. double), whereas negative indicates it is slower (e.g. minus one indicates it is twice as slow). It can be seen that enabling vectorisation for FP64 delivers marginal benefit and whilst some benefit of FP64 vectorisation with the \emph{basic} class can be observed, this is mainly due to one kernel which operates on integers driving the average upwards.

From Figure \ref{fig:vect-scalar} it can be observed that there is a greater benefit in enabling vectorisation for single precision, FP32. This benefit does vary considerably across the different kernels in each class, and hence the average for each class is fairly low, but one can see from the whiskers on the plot that there are some kernels that strongly benefit from vectorisation such as those in the \emph{stream} class. Whilst some individual kernels ran slower with vectorisation enabled, for FP32 these are in the minority and the benefits across the suite outweigh them. There are more kernels that run slower with vectorisation for FP64, however as can be seen from the whiskers in Figure \ref{fig:vect-scalar} the overhead of even the worst performing kernels tends to be small. Therefore from these results it is our recommendation that vectorisation should be enabled where possible when compiling for the SG2042.

In Section \ref{sec:riscv-comparison} it was hypothesised that one reason behind the difference in RVV FP32 and FP64 performance could be due to the 128 bit vector registers. However, with a very simple vector addition example using intrinsics it was discovered vectorisation for both single and double precision was effective in comparison to the scalar workloads. This, in part, also suggests that potentially the compiler is having an impact here, where potentially automatic vectorisation with FP64 is less effective than FP32. Consequently, an important consideration is whether to use GCC or Clang when compiling vectorised kernels for the SG2042. Indeed, Clang is often able to automatically vectorise a wider variety of code \cite{feng2021evaluation} for RISC-V then GCC, where \cite{lee2023test} found that out of the 64 kernels in the RAJAPerf benchmark suite only 30 were auto-vectorised by GCC and out of those 30 the scalar code path was executed for 7 of these at runtime. By comparison, Clang was able to auto-vectorise 59 kernels with only 3 of these following the scalar path at runtime. This is one of the reasons why the average benefit across the benchmark classes when enabling vectorisation is fairly low in Figure \ref{fig:vect-scalar}, because GCC is unable to vectorise many of these individual benchmark kernels. Indeed, the \emph{stream} class is unique as GCC is able to vectorise all of its constituent kernels, and this demonstrated by far the largest average improvement when enabling vectorisation. 

It is not only because of this improved ability to auto-vectorise that it would be useful to compile with Clang, but furthermore, in addition to supporting Vector Length Specific (VLS) RVV assembly generation, Clang can also generate Vector Length Agnostic (VLA). GCC by comparison only generates VLS RVV assembly, and therefore Clang provides more flexibility around how one might leverage the hardware and it is also interesting to explore the performance differences that selecting one of VLA or VLS will make.

However, the Clang compiler only supports RVV v1.0, whereas the SG2042's C920 cores provide RVV v0.7.1 only which is incompatible. Consequently, it is not possible to use Clang directly to compile code targeting the C920's RVV because the assembly will be incompatible. To enable experimentation with Clang we leveraged the RVV-rollback \cite{lee2023test} tool which operates upon RVV v1.0 assembly and rewrites it to backport it to RVV v0.7.1. Whilst this tool is experimental, and we would not expect users to use it for their production HPC codes on the SG2042, it does enable us to explore the use of Clang, and the different configurations it provides, for our benchmark on the CPU.

\begin{figure}[htb]
\centering
 \includegraphics[width=\columnwidth]{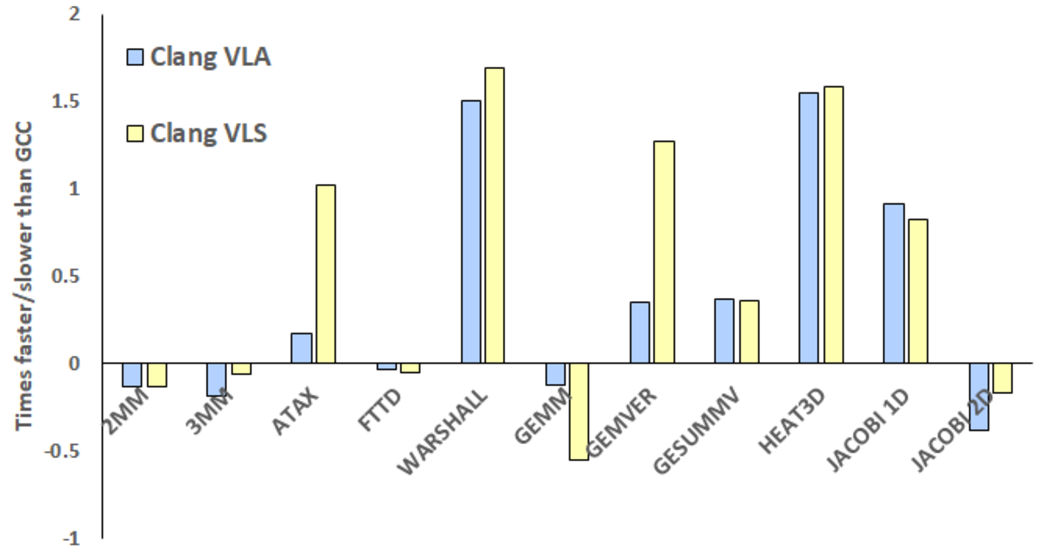}
\caption{Clang VLA and VLS single core comparison against using GCC for selected Polybench kernels in FP32}	
\label{fig:clang}
\end{figure}

Figure \ref{fig:clang} reports a single core performance comparison when using Clang in VLA and VLS mode baselined against GCC8.3 for selected Polybench kernels in FP32. Zero is the same performance, a positive number meaning Clang VLA or VLS is faster, and a negative number slower. Out of these kernels it should be noted that GCC is unable to auto-vectorise the \emph{Warshall} and \emph{Heat3D} kernels, and furthermore whilst \emph{Jacobi1D} and \emph{Jacobi2D} are vectorised by GCC the scalar code path is chosen for execution at runtime. 

By contrast, Clang is able to vectorise all the kernels but the \emph{2MM}, \emph{3MM} and \emph{GEMM} kernels execute in scalar mode only and switching to Clang delivers worse performance for these three kernels. For most of the other kernels there is a benefit to compiling with Clang, and VLS tends to outperform VLA mode. However, a surprise was that the \emph{Jacobi2D} kernel is slower with Clang compared to its GCC counterpart, which is contrary to the findings of \cite{lee2023test} however that study was running on the Allwinner D1's far simpler C906 core.

We deduce from this experiment that VLS tends to outperform VLA on the C920 and it is unfortunate that Clang only supports RVV v1.0 as using this compiler would likely be beneficial for many codes running on the SG2042. However, for optimal performance it is desirable to experiment with both Clang and GCC, on a kernel by kernel basis. 

\subsection{x86 performance comparison}
\label{sec:x86_performance}
Until this point we have explored performance of the SG2042 against other commodity RISC-V CPUs, and explored the SG2042's multi-threading and vectorisation behaviour. Our experiments have demonstrated that this CPU significantly outperforms existing widely available RISC-V hardware, but for HPC workloads that is not in-fact the competition. Instead, the critical question that must be answered for the HPC community is how the SG2042 compares against other CPUs that are currently in use in current generation supercomputers. Consequently, in this section we explore performance of the SG2042 against the x86 CPUs that are summarised in Table \ref{tab:x86-systems}. We only execute on physical cores of these x86 CPUs as all SMT is disabled by default.

The AMD Rome EPYC 7742 CPU is found in ARCHER2, which is a Cray EX and the UK national supercomputer. Similar to the SG2042, the EPYC 7742 contains 64 physical cores across four NUMA regions, each with 16 cores, but has eight instead of four memory controllers. Each core contains 32KB of I and 32KB of D L1 cache, 512 KB of L2 cache, and there is 16MB of L3 cache shared between four cores. Providing AVX2, the EPYC 7742 has 256-bit wide vector registers, which is double that of the SG2042. 

\begin{table}[htb]
\begin{center}
\begin{tabular}{ |ccccc| } 
 \hline
 CPU & Part & Clock & Cores & Vector\\ \hline
 AMD Rome & EPYC 7742 & 2.25GHz & 64 & AVX2 \\ 
 Intel Broadwell & Xeon E5-2695 & 2.1GHz & 18 & AVX2 \\ 
 Intel Icelake & Xeon 6330 & 2.0GHz & 28 & AVX512 \\ 
 Intel Sandybridge & Xeon E5-2609 & 2.40GHz & 4 & AVX \\ 
 \hline
\end{tabular}
\caption{Summary of x86 CPUs used to compare against the SG2042}
\label{tab:x86-systems}
\end{center}
\end{table}

The Intel Broadwell CPU is in Cirrus, an SGI/HPE 8600 Cluster, and the 18 physical cores reside in one NUMA region. This Xeon E5-2695 provides 32KB of I and 32KB of D L1 cache, 256KB of L2 cache which is on average the same per-core as the SG2042, and 45MB of L3 cache shared across the cores. Similarly to the AMD Rome CPU, the Xeon E5-2695 supports AVX2 and there are four memory controllers. The Intel Icelake CPU is the newest CPU that we compare against and all 28 physical cores are in a single NUMA region with 8 memory controllers. This Xeon 6330 has 32KB I and 48KB D L1 cache, 1MB L2 cache per core, which is four times that of the SG2042, and 43MB shared L3 cache. Providing AVX512, the Xeon 6330 provides 512-bit wide vector registers.

The Intel Sandybridge is the oldest CPU we compared against and the only x86 CPU we consider that is not being actively used in a current generation supercomputer. Released in 2012, it is interesting to explore how the performance of a decade old x86 CPU compares against the SG2042. Providing four physical cores, each one has 64KB of I and 64KB of D L1 cache, as well as 256KB of L2 cache and shared 10MB L3 cache. This E5-2609 supports AVX and therefore the vector register lengths are the same, 128-bit, as the SG2042. 

In all experiments conducted in this section we bind to physical cores of the x86 system and disable hyperthreading. We use GCC version 8.3 on all systems apart from ARCHER2, where GCC version 11.2 is used because that is the nearest available version. Compilation is undertaken at optimisation level O3 throughout. On all systems we execute over the most performant number of threads, on all the x86 systems this was found to be the same as the number of physical cores, whereas for the SG2042 it was demonstrated in Section \ref{sec:perf_exploration} that for some benchmark classes 32 threads provided better performance compared to 64 threads. 

\begin{figure}[htb]
\centering
 \includegraphics[width=\columnwidth]{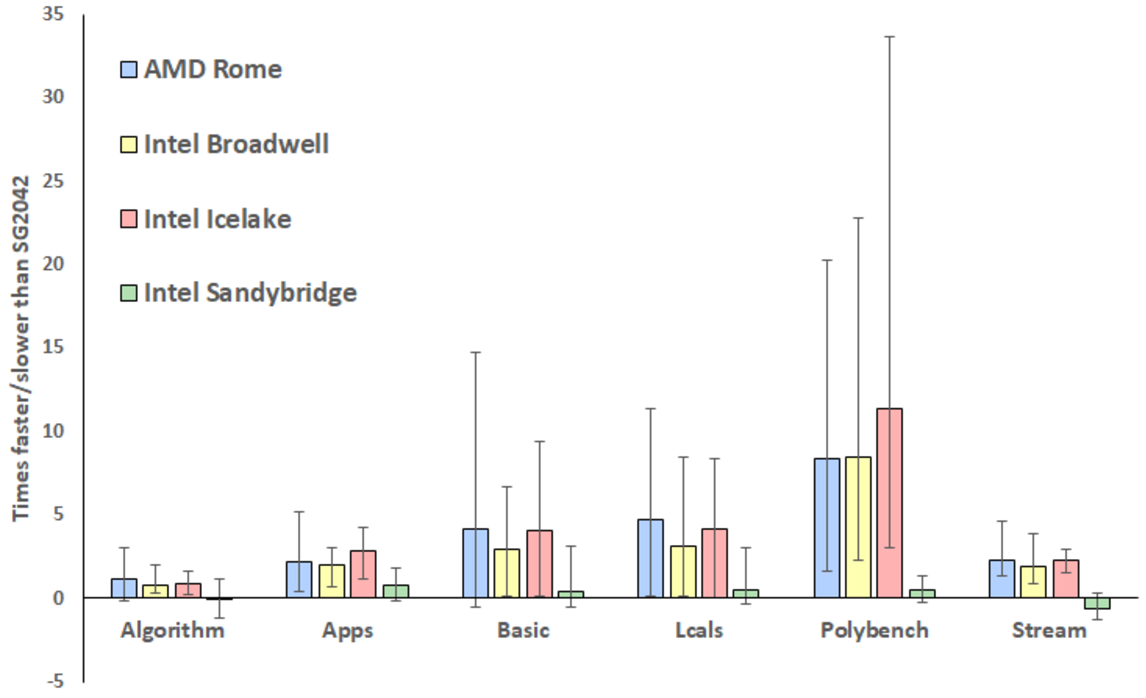}
\caption{FP64 single core comparison against x86, reporting number of times faster or slower than the baseline SG2042}	
\label{fig:threading_single_core_double}
\end{figure}

\begin{figure}[htb]
\centering
 \includegraphics[width=\columnwidth]{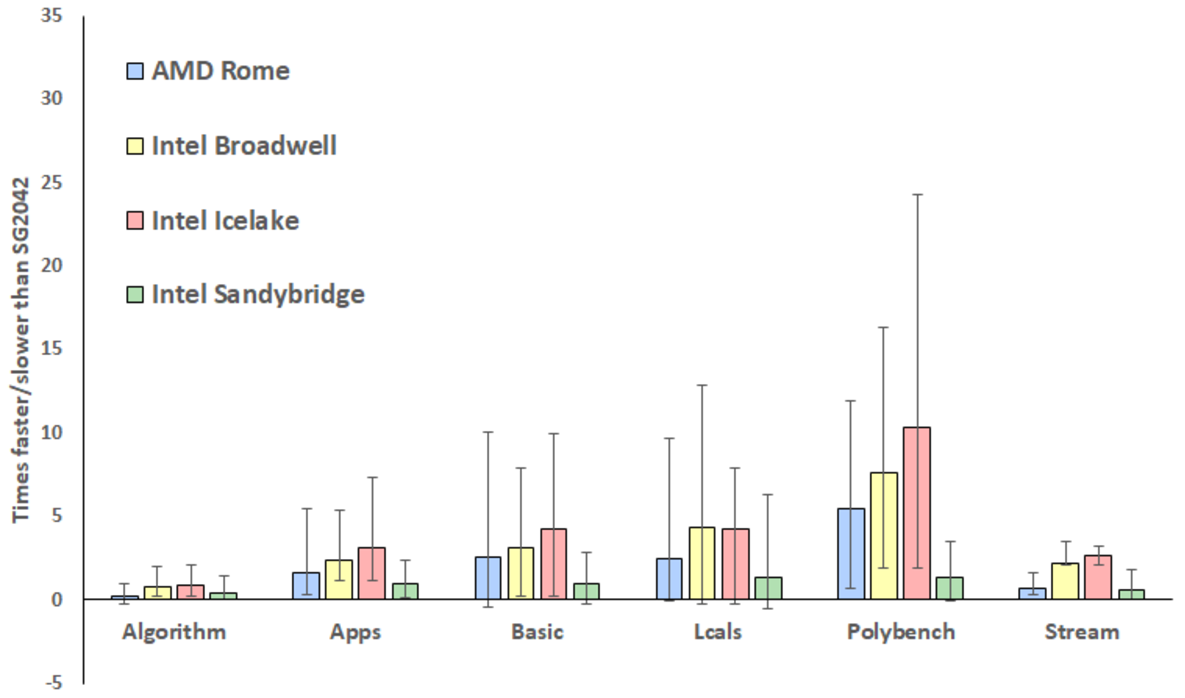}
\caption{FP32 single core comparison against x86, reporting number of times faster or slower than the baseline SG2042}	
\label{fig:threading_single_core_single}
\end{figure}

Figure \ref{fig:threading_single_core_double} reports single core performance running the benchmark suite at FP64 for the x86 CPUs baselined against the SG2042. This graph is organised the same way as the RISC-V commodity hardware comparison graph, where the bars are the average number of times faster or slower across the class, and whiskers range from the maximum to the minimum. It can be seen that all x86 cores tend to outperform the C920 apart from the Sandybridge core which on average performs slower for \emph{stream} and \emph{algorithm} benchmark classes. The AMD Rome and Intel Icelake CPUs tend to outperform the Intel Broadwell, which is understandable given that the Broadwell is the older of the three. Figure \ref{fig:threading_single_core_single} reports results from the same experiment using FP32 and it can be seen that the AMD Rome CPU is fairly lacklustre when executing at single precision compared to double, whereas the Intel processors on average perform just as well, and indeed the Sandybridge outperforms the C920 on average in each class when using FP32.

However, the average bars in Figure \ref{fig:threading_single_core_single} do not provide a complete picture. As described in Section \ref{sec:perf_exploration}, FP32 tends to perform considerably better than FP64 with RVV on the C920, and indeed it can be seen from the whiskers in Figures \ref{fig:threading_single_core_single} and \ref{fig:threading_single_core_double} that the maximum times faster is less for many benchmarks classes at FP32 than FP64. Furthermore, there are more slowest running kernels that perform slower on the x86 CPUs than the C920 at FP32. These kernels are where auto-vectorisation is being applied effectively, and indeed it can be seen that for the \emph{lcals} benchmark class there is at-least one kernel on all the x86 CPUs that performs slower than the C920. 

\begin{figure}[htb]
\centering
 \includegraphics[width=\columnwidth]{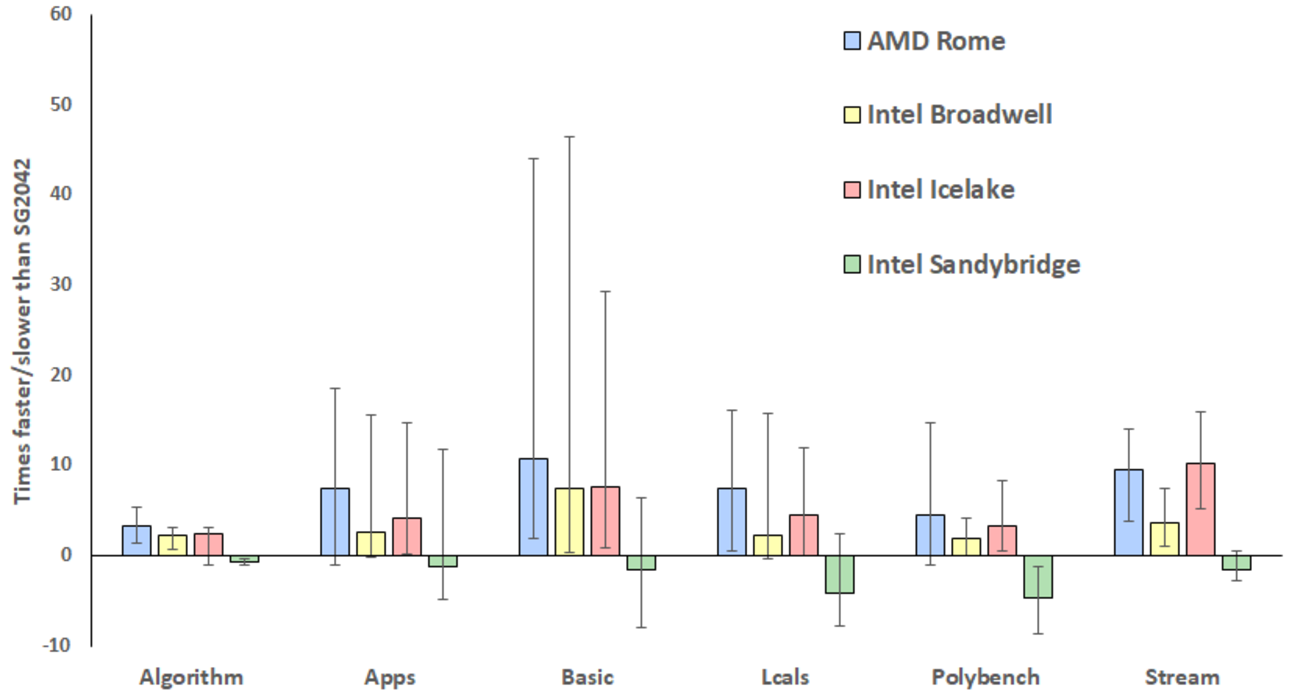}
\caption{FP64 multithreaded comparison against x86, reporting number of times faster or slower than the baseline SG2042}	
\label{fig:threading_multi_core_double}
\end{figure}

To this point we have explored single-core performance between the SG2042 and x86 CPUs. However, the SG2042 contains 64 cores which is an impressive number and only matched by the AMD Rome CPU. It is therefore instructive to undertake a performance comparison when multi-threading to understand the total performance that each CPU can deliver. Figure \ref{fig:threading_multi_core_double} reports a performance comparison of the x86 CPUs against the SG2042 for double precision, FP64. It can be seen that the \emph{basic}, \emph{lcals}, \emph{polybench}, and \emph{stream} classes benefit most from a greater number of cores and for these the SG2042 on average outperforms the Intel Sandybridge CPU. The AMD Rome, Intel Broadwell, and Intel Icelake on average outperform the SG2042, and in many cases by a significant amount.

\begin{figure}[htb]
\centering
 \includegraphics[width=\columnwidth]{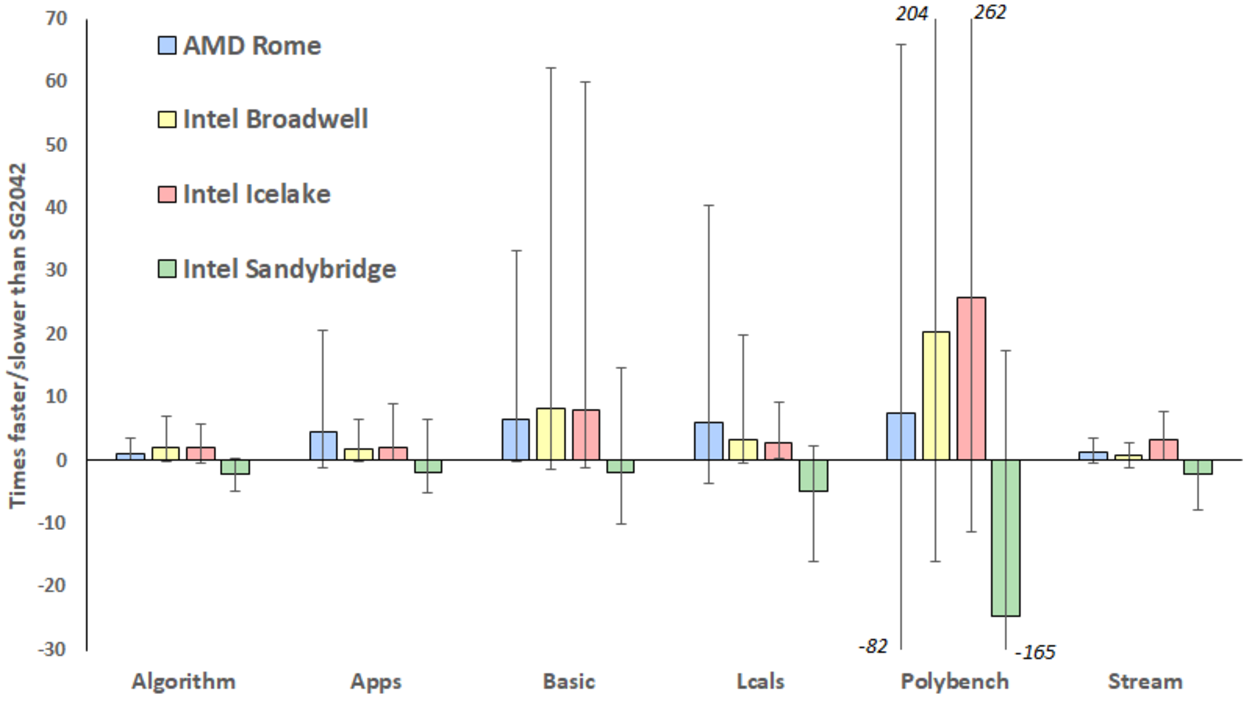}
\caption{FP32 multithreaded comparison against x86, reporting number of times faster or slower than the baseline SG2042}	
\label{fig:threading_multi_core_single}
\end{figure}

Figure \ref{fig:threading_multi_core_single} reports the multi-threaded performance comparison for FP32 and these results contain the largest variance. To aid in readability we have limited the vertical axis and labelled whisker vales that exceed this. The SG2042 tends to perform marginally more competitive against the x86 CPUs for multi-threaded FP32 than FP64, although the \emph{polybench} class is an anomaly as it performs much better on the three newest x86 CPUs than the SG2042 and on the Intel Sandybridge considerably worse on average. 

\section{Conclusions and further work}
\label{sec:conc}
In this paper we have undertaken an independent performance exploration of the Sophon SG2042 CPU. As the world's first widely available large core-count RISC-V CPU that is advertised as targeting high performance workloads, this processor could potentially drive significantly increased interest in, and adoption of, RISC-V by the HPC community. However, a critical question is whether the SG2042 actually delivers the performance that it promises and how this compares against the existing x86 CPUs that are ubiquitous in current generation supercomputers.

We demonstrated that the XuanTie C920 core outperforms the U75, which is present in the SiFive VisionFive V2, on average between five and ten times for single precision and between three and six times for double precision workloads. Based upon further exploration it was found that vectorisation is most effective with FP32 compared to FP64, and hence explaining the difference in performance between the C920 and U75 for double and single precision. Furthermore, we found that Vector Length Specific (VLS) mode tends to perform better than Vector Length Agnostic (VLA) although the difference between whether to use GCC and Clang varies on a kernel by kernel basis and is largely influenced by which tool can most effectively auto-vectorise the code and execute the vector code path.

The SG2042 contains 64 cores and we explored speed up and parallel efficiency when threading over these. In terms of programmer best practice, it was found that it is important to map threads to cores with the architecture in mind, and doing this cyclically across both NUMA regions and clusters of four cores tended to be optimal, especially up to and including 32 thread, because one is distributing across each of the four memory controllers, one per NUMA region, and the 1MB of shared L2 cache per cluster. 

We compared performance against four high performance x86 CPUs, three of which are common in current generation supercomputers and an older processor which was found in the previous generation. It was discovered that, in the main, the high performance x86 CPUs outperformed the SG2042. For a single core comparison, on average the AMD Rome performed three times faster, the Broadwell four times, the Ice Lake four times and the Sandybridge twice for FP32, these numbers were four times faster, four times, five times, and 20\% faster respectively for FP64.

When comparing multi-threaded performance against the x86 CPUs, the 64 cores of the SG2042 outperformed the 4 cores of the Sandybride on average across all the benchmark classes running at both FP32 and FP64. The equal core count and more powerful individual cores of the AMD Rome meant that this processor outperformed the SG2042 by between eight and five times for FP32 and FP64 respectively. Even though it provided a lower core count, the 18-core Broadwell outperformed the SG2042 on average six and four times for single and double precision respectively. Lastly, the 28-core IceLake, which was the newest CPU that we compared against, outperformed the SG2042 by six and eight times for FP32 and FP64. The fact that the greatest performance difference between the x86 CPUs was with single precision surprised us as we had found that the SG2042's C920 delivers better performance for FP32 compared to FP64. However clearly the x86 CPUs, especially the Intel Xeons, are also making good use of the reduced precision.

For further work we believe that it would be instructive to explore distributed memory performance on systems built around the SG2042, especially the performance that can be delivered using MPI. As the SG2042 continues to become more widely available then clusters of networked machines containing this processor will become available and would be an ideal system to undertake such experiments upon. Whilst networking performance would also be driven by the auxiliaries coupled with the CPU, not least the network adaptor, understanding what the options are in this regard would be beneficial in taking a more general view of whether machines built using the SG2042, and future CPUs in this family, would be capable of executing large scale HPC workloads.

We conclude that the SG2042 is a very exciting RISC-V technology and provides a significant step change over currently available commodity RISC-V hardware. Whilst performance is yet to match that of x86, it should be highlighted that the RISC-V vendors have come an extremely long way in a short time, and by contrast the x86 CPUs that we tested against have a very long heritage and benefit from a great many more person years of effort in their development. As T-Head produces new high performance cores and OEMs integrate these into new CPUs, for the next generation of high performance RISC-V processors it would be very useful to have RVV v1.0 provided as this will deliver the opportunity to use mainline GCC and Clang for compiling vectorised code. Furthermore, wider vector registers, increased L1 cache, and more memory controllers per NUMA region would also likely deliver significant performance advantages and help close the gap with x86 high performance processors.


\begin{acks}
This work has been funded by the ExCALIBUR H\&ES RISC-V testbed. This work used the ARCHER2 UK National Supercomputing Service (https://www.archer2.ac.uk). This work used the Cirrus UK National Tier-2 HPC Service at EPCC (http://www.cirrus.ac.uk) funded by the University of Edinburgh and EPSRC (EP/P020267/1). We thank PerfXLab for access to the SG2042 used in this work.
\end{acks}

\bibliographystyle{ACM-Reference-Format}
\bibliography{sample-base}


\begin{thebibliography}{16}


\ifx \showCODEN    \undefined \def \showCODEN     #1{\unskip}     \fi
\ifx \showDOI      \undefined \def \showDOI       #1{#1}\fi
\ifx \showISBNx    \undefined \def \showISBNx     #1{\unskip}     \fi
\ifx \showISBNxiii \undefined \def \showISBNxiii  #1{\unskip}     \fi
\ifx \showISSN     \undefined \def \showISSN      #1{\unskip}     \fi
\ifx \showLCCN     \undefined \def \showLCCN      #1{\unskip}     \fi
\ifx \shownote     \undefined \def \shownote      #1{#1}          \fi
\ifx \showarticletitle \undefined \def \showarticletitle #1{#1}   \fi
\ifx \showURL      \undefined \def \showURL       {\relax}        \fi
\providecommand\bibfield[2]{#2}
\providecommand\bibinfo[2]{#2}
\providecommand\natexlab[1]{#1}
\providecommand\showeprint[2][]{arXiv:#2}

\bibitem[Beckingsale et~al\mbox{.}(2019)]%
        {beckingsale2019raja}
\bibfield{author}{\bibinfo{person}{David~A Beckingsale}, \bibinfo{person}{Jason Burmark}, \bibinfo{person}{Rich Hornung}, \bibinfo{person}{Holger Jones}, \bibinfo{person}{William Killian}, \bibinfo{person}{Adam~J Kunen}, \bibinfo{person}{Olga Pearce}, \bibinfo{person}{Peter Robinson}, \bibinfo{person}{Brian~S Ryujin}, {and} \bibinfo{person}{Thomas~RW Scogland}.} \bibinfo{year}{2019}\natexlab{}.
\newblock \showarticletitle{RAJA: Portable performance for large-scale scientific applications}. In \bibinfo{booktitle}{\emph{2019 ieee/acm international workshop on performance, portability and productivity in hpc (p3hpc)}}. IEEE, \bibinfo{pages}{71--81}.
\newblock


\bibitem[eProcessor project(2023)]%
        {eprocessor}
eProcessor project \bibinfo{year}{2023}\natexlab{}.
\newblock \bibinfo{booktitle}{\emph{eProcessor: an open source full stack ecosystem}}.
\newblock
\urldef\tempurl%
\url{https://eprocessor.eu/}
\showURL{%
Retrieved Aug 16, 2023 from \tempurl}


\bibitem[Esperanto Technologies(2023)]%
        {esperanto}
Esperanto Technologies \bibinfo{year}{2023}\natexlab{}.
\newblock \bibinfo{booktitle}{\emph{Esperanto: Outstanding solutions for Generative AI and HPC}}.
\newblock
\urldef\tempurl%
\url{https://www.esperanto.ai/}
\showURL{%
Retrieved Aug 16, 2023 from \tempurl}


\bibitem[Feng et~al\mbox{.}(2021)]%
        {feng2021evaluation}
\bibfield{author}{\bibinfo{person}{Jing~Ge Feng}, \bibinfo{person}{Ye~Ping He}, {and} \bibinfo{person}{Qiu~Ming Tao}.} \bibinfo{year}{2021}\natexlab{}.
\newblock \showarticletitle{Evaluation of compilers’ capability of automatic vectorization based on source code analysis}.
\newblock \bibinfo{journal}{\emph{Scientific Programming}}  \bibinfo{volume}{2021} (\bibinfo{year}{2021}), \bibinfo{pages}{1--15}.
\newblock


\bibitem[Herdt et~al\mbox{.}(2020)]%
        {herdt2020risc}
\bibfield{author}{\bibinfo{person}{Vladimir Herdt}, \bibinfo{person}{Daniel Gro{\ss}e}, \bibinfo{person}{Pascal Pieper}, {and} \bibinfo{person}{Rolf Drechsler}.} \bibinfo{year}{2020}\natexlab{}.
\newblock \showarticletitle{RISC-V based virtual prototype: An extensible and configurable platform for the system-level}.
\newblock \bibinfo{journal}{\emph{Journal of Systems Architecture}}  \bibinfo{volume}{109} (\bibinfo{year}{2020}), \bibinfo{pages}{101756}.
\newblock


\bibitem[Hornung and Hones(2017)]%
        {hornung2017raja}
\bibfield{author}{\bibinfo{person}{Richard~D Hornung} {and} \bibinfo{person}{Holger~E Hones}.} \bibinfo{year}{2017}\natexlab{}.
\newblock \bibinfo{booktitle}{\emph{Raja performance suite}}.
\newblock \bibinfo{type}{{T}echnical {R}eport}. \bibinfo{institution}{Lawrence Livermore National Lab.(LLNL), Livermore, CA (United States)}.
\newblock


\bibitem[Jesus({[n.\,d.]})]%
        {jesus2023check}
\bibfield{author}{\bibinfo{person}{Ricardo Jesus}.} \bibinfo{year}{[n.\,d.]}\natexlab{}.
\newblock \showarticletitle{Check for A Study on the Performance Implications of AArch64 Atomics Ricardo Jesus () and Mich{\`e}le Weiland EPCC, The University of Edinburgh, Edinburgh, UK}. In \bibinfo{booktitle}{\emph{High Performance Computing: 38th International Conference, ISC High Performance 2023, Hamburg, Germany, May 21--25, 2023, Proceedings}}. Springer Nature, \bibinfo{pages}{279}.
\newblock


\bibitem[Jesus and Weiland(2022)]%
        {jesus2022chapelperf}
\bibfield{author}{\bibinfo{person}{Ricardo Jesus} {and} \bibinfo{person}{Mich{\`e}le Weiland}.} \bibinfo{year}{2022}\natexlab{}.
\newblock \showarticletitle{ChapelPerf: A Performance Suite for Chapel}. In \bibinfo{booktitle}{\emph{The Annual Chapel Implementers and Users Workshop}}.
\newblock


\bibitem[Jost et~al\mbox{.}(2021)]%
        {jost2021seamless}
\bibfield{author}{\bibinfo{person}{Tiago~Trevisan Jost}, \bibinfo{person}{Yves Durand}, \bibinfo{person}{Christian Fabre}, \bibinfo{person}{Albert Cohen}, {and} \bibinfo{person}{Fr{\'e}d{\'e}ric P{\'e}rrot}.} \bibinfo{year}{2021}\natexlab{}.
\newblock \showarticletitle{Seamless compiler integration of variable precision floating-point arithmetic}. In \bibinfo{booktitle}{\emph{2021 IEEE/ACM International Symposium on Code Generation and Optimization (CGO)}}. IEEE, \bibinfo{pages}{65--76}.
\newblock


\bibitem[Lee et~al\mbox{.}(2023a)]%
        {lee2023backporting}
\bibfield{author}{\bibinfo{person}{Joseph~KL Lee}, \bibinfo{person}{Maurice Jamieson}, {and} \bibinfo{person}{Nick Brown}.} \bibinfo{year}{2023}\natexlab{a}.
\newblock \showarticletitle{Backporting risc-v vector assembly}.
\newblock \bibinfo{journal}{\emph{arXiv preprint arXiv:2304.10324}} (\bibinfo{year}{2023}).
\newblock


\bibitem[Lee et~al\mbox{.}(2023b)]%
        {lee2023test}
\bibfield{author}{\bibinfo{person}{Joseph~KL Lee}, \bibinfo{person}{Maurice Jamieson}, \bibinfo{person}{Nick Brown}, {and} \bibinfo{person}{Ricardo Jesus}.} \bibinfo{year}{2023}\natexlab{b}.
\newblock \showarticletitle{Test-driving RISC-V Vector hardware for HPC}.
\newblock \bibinfo{journal}{\emph{arXiv preprint arXiv:2304.10319}} (\bibinfo{year}{2023}).
\newblock


\bibitem[Mantovani et~al\mbox{.}(2023)]%
        {mantovani2023software}
\bibfield{author}{\bibinfo{person}{Filippo Mantovani}, \bibinfo{person}{Pablo Vizcaino}, \bibinfo{person}{Fabio Banchelli}, \bibinfo{person}{Marta Garcia-Gasulla}, \bibinfo{person}{Roger Ferrer}, \bibinfo{person}{Giorgos Ieronymakis}, \bibinfo{person}{Nikos Dimou}, \bibinfo{person}{Vassilis Papaefstathiou}, {and} \bibinfo{person}{Jesus Labarta}.} \bibinfo{year}{2023}\natexlab{}.
\newblock \showarticletitle{Software Development Vehicles to enable extended and early co-design: a RISC-V and HPC case of study}.
\newblock \bibinfo{journal}{\emph{arXiv preprint arXiv:2306.01797}} (\bibinfo{year}{2023}).
\newblock


\bibitem[Open chip community(2023)]%
        {c906}
Open chip community \bibinfo{year}{2023}\natexlab{}.
\newblock \bibinfo{booktitle}{\emph{Open XuanTie C906}}.
\newblock
\urldef\tempurl%
\url{https://xrvm.com/cpu-details?id=4056751997003636736}
\showURL{%
Retrieved Aug 16, 2023 from \tempurl}


\bibitem[Perez et~al\mbox{.}(2021)]%
        {perez2021coyote}
\bibfield{author}{\bibinfo{person}{Boria Perez}, \bibinfo{person}{Alexander Fell}, {and} \bibinfo{person}{John~D Davis}.} \bibinfo{year}{2021}\natexlab{}.
\newblock \showarticletitle{Coyote: An open source simulation tool to enable RISC-V in HPC}. In \bibinfo{booktitle}{\emph{2021 Design, Automation \& Test in Europe Conference \& Exhibition (DATE)}}. IEEE, \bibinfo{pages}{130--135}.
\newblock


\bibitem[rvv-next(2023)]%
        {rvv-next}
rvv-next \bibinfo{year}{2023}\natexlab{}.
\newblock \bibinfo{booktitle}{\emph{GNU compiler collection}}.
\newblock
\urldef\tempurl%
\url{https://github.com/riscv-collab/riscv-gnu-toolchain/tree/rvv-next}
\showURL{%
Retrieved Aug 16, 2023 from \tempurl}


\bibitem[StarFive(2023)]%
        {starfive-soc}
StarFive \bibinfo{year}{2023}\natexlab{}.
\newblock \bibinfo{booktitle}{\emph{SoC Platform}}.
\newblock
\urldef\tempurl%
\url{https://www.starfivetech.com/en/site/soc}
\showURL{%
Retrieved Aug 16, 2023 from \tempurl}


\end{thebibliography}


\end{document}